\documentclass[layout=onecolumn]{achemso}

\usepackage[latin9]{inputenc}
\usepackage[T1]{fontenc}
\usepackage[english]{babel}

\usepackage{algorithm}
\usepackage{amsmath}
\usepackage{amssymb}
\usepackage{color}
\usepackage{graphicx}
\usepackage{listings}

\newcommand*{\N}{\ensuremath{\mathbb{N}}}
\newcommand*{\R}{\ensuremath{\mathbb{R}}}
\newcommand*{\C}{\ensuremath{\mathbb{C}}}
\newcommand*{\e}{\ensuremath{\varepsilon}}

\DeclareMathOperator*{\argmax}{arg\,max}
\DeclareMathOperator*{\argmin}{arg\,min}

\SectionNumbersOn
\setkeys{acs}{usetitle=true}

\makeatletter
\title{Iterative Power Algorithm for Global Optimization with Quantics Tensor Trains}

\author{Micheline B. Soley}
\affiliation{Yale Quantum Institute, Yale University, P.O. Box 208334, New Haven, CT, 06520-8263, USA}
\alsoaffiliation{Department of Chemistry, Yale University, P.O. Box 208107, New Haven, CT, 06520, USA}
\author{Paul Bergold}
\affiliation{Zentrum Mathematik, Technical University of Munich, Boltzmannstr. 3, 85748 Garching, Germany}
\author{Victor S. Batista}
\affiliation{Yale Quantum Institute, Yale University, P.O. Box 208334, New Haven, CT, 06520-8263, USA}
\alsoaffiliation{Department of Chemistry, Yale University, P.O. Box 208107, New Haven, CT, 06520, USA}
\email{victor.batista@yale.edu}

\makeatother
\begin{document}
\newpage
\begin{abstract}
Optimization algorithms play a central role in chemistry since optimization is the computational keystone of most molecular and electronic structure calculations. Herein, we introduce the iterative power algorithm (IPA) for global optimization and a formal proof of convergence for both discrete and continuous global search problems, which is essential for applications in chemistry such as molecular geometry optimization. IPA implements the power iteration method in quantics tensor train (QTT) representations. Analogous to the imaginary time propagation method with infinite mass, IPA starts with an initial probability distribution $\rho_0({\bf x})$ and iteratively applies the recurrence relation $\rho_{k+1}({\bf x})=U({\bf x})\rho_k({\bf x})/\|U\rho_k\|_{L^1}$, where $U({\bf x})=e^{-V({\bf x})}$ is defined in terms of the potential energy surface (PES) $V({\bf x})$ with global minimum at ${\bf x}={\bf x}^*$. Upon convergence, the probability distribution becomes a delta function $\delta({\bf x}-{\bf x}^*)$, so the global minimum can be obtained as the position expectation value ${\bf x}^* = \text{Tr}\left[{\bf x}\,\delta({\bf x}-{\bf x}^*)\right]$. QTT representations of $V({\bf x})$ and $\rho({\bf x})$ are generated by fast adaptive interpolation of multidimensional arrays to bypass the curse of dimensionality and the need to evaluate $V({\bf x})$ for all possible values of ${\bf x}$. We illustrate the capabilities of IPA for global search optimization of two multidimensional PESs, including a differentiable model PES of a DNA chain with $D=50$ adenine-thymine base pairs, and a discrete non-differentiable potential energy surface, $V(p)=\text{mod}(N,p)$, that resolves the prime factors of an integer $N$, with $p$ in the space of prime numbers $\{2,3,\dots,p_{\max}\}$ folded as a $d$-dimensional $2_1 \times 2_2 \times \cdots \times 2_d$ tensor. We find that IPA resolves multiple degenerate global minima even when separated by large energy barriers in the highly rugged landscape of the potentials. Therefore, IPA should be of great interest for a wide range of other optimization problems ubiquitous in molecular and electronic structure calculations.
\end{abstract}

\newpage
\section{Introduction}
The development of efficient optimization algorithms remains a subject of great research interest in chemistry and beyond since optimization is essential for most molecular and electronic structure calculations. In control of chemical processes, for example, global optimization algorithms are essential to determine the drives that steer a system into a desired final state.\cite{Bellman.1961,Li.2018.033602,Shi.1988.6870,Shi.1989.185,Peirce.1988.4950,Kosloff.1989.201,Jakubetz.1990.100,Rego.2009.293,Brif.2010.07008,Soley.2015.715,Videla.2018.1198,Soley.2018.3351} 
Another prototypical example is the problem of finding the minimum energy structure of a complex molecule, usually the first step in studies of molecular properties, molecular reactivity, and drug design.\cite{Levinthal.1969.22,Sali.1994.248,Wales.2000.1,Dill.2008.289}
The simplest approach for finding the global optima in a discrete set is to sift through all possibilities. However, that approach becomes intractable for high-dimensional systems since the number of possible states typically scales exponentially with the number of degrees of freedom -- {\em i.e.}, the so-called ``curse of dimensionality'' problem.\cite{Bellman.1961}
Analogously, simple approaches for continuous optimization involve sampling stochastically\cite{Fogel.1962.14,Pincus.1968.690,Cavicchio.1970.Adaptive,Pincus.1970.1225,Holland.1975,Kirkpatrick.1983.621,Cerny.1985.41,Li.1987.6611,Koza.1989.768,Koza.1990.Genetic,Wales.1997.5111}
or deterministically.\cite{1961.Hooke,1962.Spendley,1965.Nelder,Land.1960.497,Little.1963.972,Glover.1985.557,Glover.1985.Future,1993.Amara,1996.Andricioaei,Piela.1989.3339,Pillardy.1992.4337,1963.Fletcher,1964.Fletcher,1964.Lee,1967.Broyden,1970.Broyden,1970.Goldfarb,1970.Shanno,1995.Byrd,2011.Morales,1997.Zhu,Soley.2015.715,Soley.2018.3351}
 Yet, these procedures typically lead to ``trapping'' in local minima. Therefore, the development of efficient global search algorithms remains an open problem of great interest.

In this paper, we build upon the strategy of the diffeomorphic modulation under observable-response-preserving homotopy (DMORPH) method,\cite{Soley.2018.3351} and we introduce the iterative power algorithm (IPA) for global optimization. DMORPH evolves a distribution function $\rho({\bf x})$ in the search space of configurations, so that the distribution becomes localized at the global optima and the global minimum position can be revealed by computing the position expectation value.\cite{Soley.2018.3351} Analogously, IPA implements the same strategy of evolving a probability distribution function although with a very different approach. Instead of implementing the DMORPH approach of iteratively optimizing control parameters of an externally applied field that localizes $\rho({\bf x})$ at the global optima, IPA applies a simple amplitude amplification scheme based on the power method known from numerical linear algebra.\cite{Muntz.1913.43,vMises.1929.58,vMises.1929.152,Chatelin.2012,Trefethen:1997aa} The resulting algorithm is essentially an imaginary time propagation\cite{Kosloff.1986,Metropolis.1949.335,Donsker.1950.551,Anderson.1975.1499,Reynolds.1982.5593} although with infinite mass. The relation between the power method\cite{Muntz.1913.43,vMises.1929.58,vMises.1929.152,Chatelin.2012} and the imaginary time propagation method\cite{Kosloff.1986,Metropolis.1949.335,Donsker.1950.551,Anderson.1975.1499,Reynolds.1982.5593} has been previously discussed,\cite{Greene.2017.4034,Lehtovaara.2007.148,Bader.2013.1304.6845v2,Shani.2017.Analysis,Schwarz.2017.176403} although it remains to be formally analyzed.

The power method is based on the recurrence relation $\rho_{k+1}({\bf x})=U({\bf x})\rho_k({\bf x})/\|U\rho_k\|_{L^1}$. In the IPA implementation, $U({\bf x})=e^{-V({\bf x})}$ is defined by the scaled potential energy surface (PES) $V({\bf x})$, and $\rho_k({\bf x})$ is the density distribution after the $k$th optimization step. Such an iterative procedure transforms any initial distribution with nonzero amplitude at the global minimum into a delta function $\rho({\bf x})=\delta({\bf x}-{\bf x}^*)$ ({\em i.e.}, the eigenvector of $U({\bf x})$ with maximum eigenvalue in the basis of Dirac delta functions). The global minimum can then be revealed, as in the DMORPH method, by computing the position expectation value ${\bf x}^* = \text{Tr}\left[{\bf x}\,\delta({\bf x}-{\bf x}^*)\right]$. 

IPA can efficiently find the global minimum of low-rank high-dimensional potential energy surfaces with possible position states ${\bf x}$ by approximating $\rho({\bf x})$ and $V({\bf x})$ in $D\ge 1$ physical dimensions in the form of quantics tensor trains (QTTs) in $n\ge 1$ reshaped dimensions.\cite{Khoromskij.2011.257,Khoromskij.2010.69,Gavrilyuk.2011.273} QTTs are a specific form of tensor trains (TTs),\cite{Oseledets.2010.70,Oseledets.2011.2295} which are of great interest and themselves a specific form of matrix product states (MPS)\cite{Ostlund.1995.3537}. For the QTT format, $2^d$-element arrays, each representing a length-$2^d$ grid in a single physical dimension, are reshaped into $2_1 \times 2_2 \times \cdots \times 2_d$ tensors $Q(i_{1},\dots,i_{d})$, where quantics refers to $q$-adic folding in which each folding dimension of the reshaped tensor is represented by $q=2$ elements.\cite{Khoromskij.2011.257}
Since they depend on $d\ge 1$ folding variables $i_k$, each of them with two possible values, they are decomposed into the outer product of tensor cores in the form of a matrix product state/tensor train as\cite{Oseledets.2010.70,Oseledets.2011.2295}
\begin{align}\label{eq:QTT}
	Q(i_{1},\dots,i_{d})
	&\approx\sum_{\alpha_{1}=1}^{r_{1}}\sum_{\alpha_{2}=1}^{r_{2}}\cdots\sum_{\alpha_{d-1}=1}^{r_{d-1}}A_{1}(1,i_{1},\alpha_{1})A_{2}(\alpha_{1},i_{2},\alpha_{2})\cdots A_{d}(\alpha_{d-1},i_{d},1),
\end{align}
where $i_1,\dots,i_d\in\{1,2\}$ and $A_{j}$ are individual order-three, rank $r_{j}$ tensor cores contracted over the auxiliary indices $\alpha_{j}$ for $j=1,\dots,d$. Results for each of the physical dimensions $D$ are incorporated via outer products to form a quantics tensor train with a total of $n=d\times D$ dimensions. The QTT format, introduced by Eq.~\eqref{eq:QTT}, reduces the cost of evaluating $Q$ over the search space of $2^{d}$ possibilities to not more than $2dr^{2}$ evaluations for the maximal rank $r=\max(r_1,\dots,r_{d-1})$.\cite{Khoromskij.2011.257} This scaling is advantageous in chemistry, as many molecular processes can be modeled by low-rank sums of double well potentials, including hydrogen bonding in DNA, protonation of water molecules, and arrangement of Zundel ions. We demonstrate the capabilities of IPA as applied to determination of the optimal configuration of protons in a DNA chain of $D=50$ adenine-thymine (A-T) base pairs with $2^{50}$ local minima corresponding to all possible protonation states. 

In addition, quantics tensor trains feature the same exponential improvement in data sparsity given by quantum computers,\cite{Savostyanov.2012.3215} which offers the possibility of developing methods like IPA that can be thought of as classical computing analogues of quantum computing algorithms.

Quantum search algorithms ({\em e.g.}, the Grover's search method\cite{Grover.1996.212}) typically initialize a uniform superposition and evolve it multiple times until a measurement of the resulting state can identify one out of $2^d$ possibilities with sufficiently high probability. Analogously, we initialize $\rho_0({\bf x})$ as a uniform distribution in the QTT format to enable sampling of the entire search space simultaneously. Iterative application of the recurrence relation amplifies the amplitude at the global minima, which yields a final density $\rho_{\operatorname{final}}({\bf x})$ localized at the global minima. We prove that the number of steps required by IPA to amplify the amplitude in the presence of a single global minimum to a probability higher than 50\% scales logarithmically with the size of the search space, which provides a valuable global search methodology alternative to well-established optimization methods\cite{Eiselt:2019aa,Bomze:2010aa,Aragon:2019aa}

The paper is organized as follows. The IPA method is introduced in Section~2, followed by the analysis of the convergence rate in Section~3 and a discussion in the perspective of existing approaches in Section~4. Computational results are presented in Section~5 and conclusions in Section~6. Appendix A presents a formal proof of IPA convergence. Appendix B analyzes the convergence rate of the power method. Python codes to reproduce the reported calculations are provided in Appendices C, D, and E.

\section{Iterative Power Algorithm Method}\label{sub:ipa_sequence}
IPA solves the optimization problem of finding the global minima of a given potential $V\colon\R^n\to\R$. For simplicity, in this section, we discuss the one-dimensional case $x\in\R$. However, we demonstrate the capabilities of IPA for global optimization of model PESs with up to $n=400$ dimensions.

To guarantee the existence of a global minimum,\cite{Aragon:2019aa} we assume $V(x)$ is continuous and coercive ({\em i.e.}, $V(x)\to +\infty$ as $|x|\to +\infty$).
Our goal is to compute the set of all minima locations of $V(x)$
\begin{align}\label{def:argmin}
	\argmin_{x\in\R}\,V(x)
	=\Big\{x^*\in\R \mid V(x)\ge V(x^*)\,\,\text{for all $x\in\R$}\Big\}.
\end{align}
Therefore, we employ a non-negative probability density function $\rho_0\colon\R\to[0,\infty)$ that is bounded and with unit norm
\begin{align}
	\|\rho_0\|_{L^1}
	=\int_\R\mathrm{d}x\,\rho_0(x)
	=1.
\end{align}
The initial density $\rho_0(x)$ is supported (nonzero) around all minima locations $x^*$ of the potential $V(x)$, so for all $r>0$, the initial density satisfies the following condition
\begin{align}\label{eq:prob_r}
	\int_{x^*-r}^{x^*+r}\mathrm{d}x\,\rho_0(x)
	>0.
\end{align}
In each IPA iteration, a transformation function $U(x)$ is applied from the left to $\rho_0(x)$ to increase the density amplitude at the global minimum positions relative to amplitudes at the remainder of the search space (in discrete space, $U(x)$ is represented as a matrix, $\rho_0(x)$ as a vector, and $U(x)\rho_0(x)$ is a matrix-vector product). The resulting product $U(x)\rho_0(x)$ is then normalized to obtain a new density $\rho_1(x)$, which is the input for the next IPA iteration. Any $U(x)$ can be used, provided it satisfies the following two conditions: (i) $U(x)$ must be a continuous and strictly positive function that is maximized at the global minima of $V(x)$, {\em i.e.},
\begin{align}\label{eq:minmax}
	\argmax_{x\in\R}\,U(x)
	=\argmin_{x\in\R}\,V(x),
\end{align}
and (ii) $U(x)$ must be integrable (we denote this by $U\in L^{1}(\R)$).

A simple example is $U(x)=e^{-\beta V(x)}$ for a fixed scaling parameter $\beta>0$. We note that Eq.~\eqref{eq:minmax} holds since the exponential is a strictly increasing function. Furthermore, the coercivity condition of the potential implies that $U(x)$ is integrable for a sufficiently fast growing potential $V(x)$ in the asymptotic region $|x|\to +\infty$.

\subsection{Evolution: Amplitude Amplification}
IPA generates a sequence of density distributions $\rho_1,\rho_2,\dots$, starting from a uniform distribution $\rho_0(x)$, as follows:\\[2mm]
\noindent\hspace*{1cm}
\textbf{for}
	$k=1,2,\dots$\\[2mm]
	\hspace*{1.5cm}
	$\eta_k
	=\|U\rho_{k-1}\|_{L^1}
	=\displaystyle\int_\R\mathrm{d}x\,U(x)\rho_{k-1}(x)$;\\[4mm]
	\hspace*{1.5cm}
	$\rho_k(x)
	=\displaystyle\frac{U(x)\rho_{k-1}(x)}{\eta_k}
	=\frac{U(x)^k\rho_0(x)}{\|U^k\rho_0\|_{L^1}}$;\\[2mm]
	\hspace*{1cm}
\textbf{end}\\[2mm]
Although this expression could be implemented with a polynomial expansion (for example, a Chebyshev or Fourier series expansion), we only employ the tensor-train cross approximation as described in Section~\ref{subsec:QTTRepresentation}. Since $U(x)$ is assumed to be continuous and integrable, we conclude it is bounded and $L^1$-normalizable ({\em i.e.}, $U\in L^{\infty}(\R)\cap L^{1}(\R)$).
In particular, this guarantees the normalization factors $\eta_k>0$ are well defined, since repeated applications of $U(x)$ remain $L^1$-normalizable ({\em i.e.}, $U^k\in L^1(\R)$ for all iterations $k\ge 1$).\cite{Folland:1999aa}

The Appendix proves that the sequence $\rho_1,\rho_2,\dots$ of densities produced by IPA converges to a kind of ``Dirac comb'' distribution ({\em i.e.}, a sum of Dirac delta functions), located at the global minima positions $x_1^*< x_2^*<...< x_s^*$ of the potential
\begin{align}\label{eq:sum_delta}
	 \rho_{\operatorname{final}}(x)
	 =\lim_{k \to \infty} \rho_k(x)
	 =\sum_{j=1}^{s} \delta(x-x_j^*).	
\end{align}
where $s\ge 1$ is the number of minima positions.
As mentioned in the Appendix, the final density $\rho_{\operatorname{final}}(x)$ can be viewed as the limit of so-called Dirac sequences.

\subsection{Resolution of Global Minima: Measurement}\label{sec:measurement}
The global minima are obtained after obtaining $\rho_{\operatorname{final}}(x)$ as follows:

(i) When $V(x)$ has a single global minimum at $x=x^*$, the minimum is obtained by computing the position expectation value with the final density $\rho_{\operatorname{final}}(x)$
\begin{align}\label{eq:expectationvalue}
	x^*
	=\langle x\rangle_{\rho_{\operatorname{final}}}
	=\int_\R\mathrm{d}x\,x\rho_{\operatorname{final}}(x).
\end{align}

(ii) When $V(x)$ has only two degenerate global minima ({\em e.g.}, as for the factorization of biprimes discussed below), we first compute the position expectation value of $\rho_{\operatorname{final}}(x)$ to obtain the average position $\overline{x}$ of the two global minima. Then, we multiply the final density by a shifted Heaviside step function
\begin{align}\label{eq:Heaviside}
	\Theta(x-\bar x)
	= 		
	\begin{cases}
		0, & \text{if $x\le\overline{x}$},\\
		1, & \text{if $x>\overline{x}$},
	\end{cases}
\end{align}
to obtain the distributions $\rho_{\operatorname{final}}(x)\Theta(x-\bar x)$ and $\rho_{\operatorname{final}}(x) (1-\Theta(x-\bar x))$, which are single delta functions resolving the two distinct minima.

(iii) When $V(x)$ has an unknown number of global minima, we first obtain $\rho_{\operatorname{final}}(x)$ using IPA. Then, we reinitialize $\rho_0 \propto \rho_{\operatorname{final}}$, such that the initial density is a Dirac comb with amplitude only at the global minima positions. The first component of the Dirac comb is isolated with a second use of IPA using a ``ramp potential'' rather than the potential $V(x)$ of the problem of interest. The ramp is usually a simple monotonically increasing function ({\em e.g.}, $\text{ramp}(x)=x$) that breaks the degeneracy of the Dirac comb $\rho_0(x)$ by amplifying the amplitude of the minimum of all minima ({\em i.e.}, $x_1^*$). Since the amplitude of the density $\rho_0(x)$ is only nonzero at global minima positions, only the Dirac delta component localized at the global minimum with the lowest position remains, and the expectation value of the position then yields the location of the first global minimum of the original potential energy surface $V(x)$. After computing $x_1^*$, we multiply $\rho_{\operatorname{final}}(x)$ by the Heaviside function $\Theta(x-x_1^*)$ introduced by Eq.~(\ref{eq:Heaviside}) and we repeat the IPA ramp process to identify the second minimum ({\em i.e.}, $x_2^*$). The scheme is then repeated until all global minima are resolved.

\subsection{QTT Representation}\label{subsec:QTTRepresentation}
IPA is not limited to a specific choice of basis set representation for $\rho({\bf x})$, $V({\bf x})$, and $U({\bf x})$. However, we employ the quantics tensor train (QTT) representation,\cite{Khoromskij.2011.257,Khoromskij.2010.69,Gavrilyuk.2011.273} generated by fast adaptive interpolation of multidimensional arrays as implemented in Oseledets' TT-Toolbox.\cite{Oseledets.2020.TT} The search space of size $2^d$ in each physical dimension is reshaped into a $d$-dimensional $2_1 \times 2_2 \times \cdots \times 2_d$ tensor in a row-major order prior to tensor-train decomposition. In $D\ge 1$ physical dimensions, the full $d\times D$-dimensional search space is represented as a list of $D$ tensors, each of which is a Kronecker product of the aforementioned search space tensor with $D-1$ one tensors of the same shape. IPA is implemented here to optimize potential energy surfaces in up to $n=d\times D=400$ dimensions, as optimization is performed for a quantics tensor train with $d\ge 1$ folding dimensions of the original search space grid and $D$ physical dimensions. Operations of functions on the resulting QTTs are then calculated according to the cross approximation,\cite{Oseledets.2010.70} which determines a low-rank representation of the tensor through evaluation of a limited number of tensor entries. Functions of QTTs such as the exponential $U({\bf x})=e^{-V({\bf x})}$ are thereby determined without resorting to additional approximations such as Taylor series expansions or Padé approximants. We represent both operators and densities as tensor trains such that quantities need never be evaluated everywhere on the search space. For example, the Heaviside function $\Theta(\bf{x}-\overline{\bf{x}})$ (Eq.~\ref{eq:Heaviside}) is represented as a $2_1 \times 2_2 \times \cdots \times 2_d$ tensor that acts directly on an analogous tensor for the density $\rho_{\operatorname{final}}(\mathbf{x})$. This reproduces the action of the Heaviside operator on the density without determination of the result at all $2^d$ points, which reduces computational expense. Integrals over position space are also performed without leaving the tensor train representation. For example, the expectation value that gives the position of the global minimum is calculated as the inner product of the tensor trains for the position and the final density. The resulting implementation bypasses the curse of dimensionality and allows for applications to high-dimensional potentials (Python scripts provided in Appendices C, D, and E).

\section{Convergence Rate Analysis}\label{sec:ConvergenceRate}
\begin{figure}
	\begin{centering}
		\includegraphics[width=0.7\textwidth]{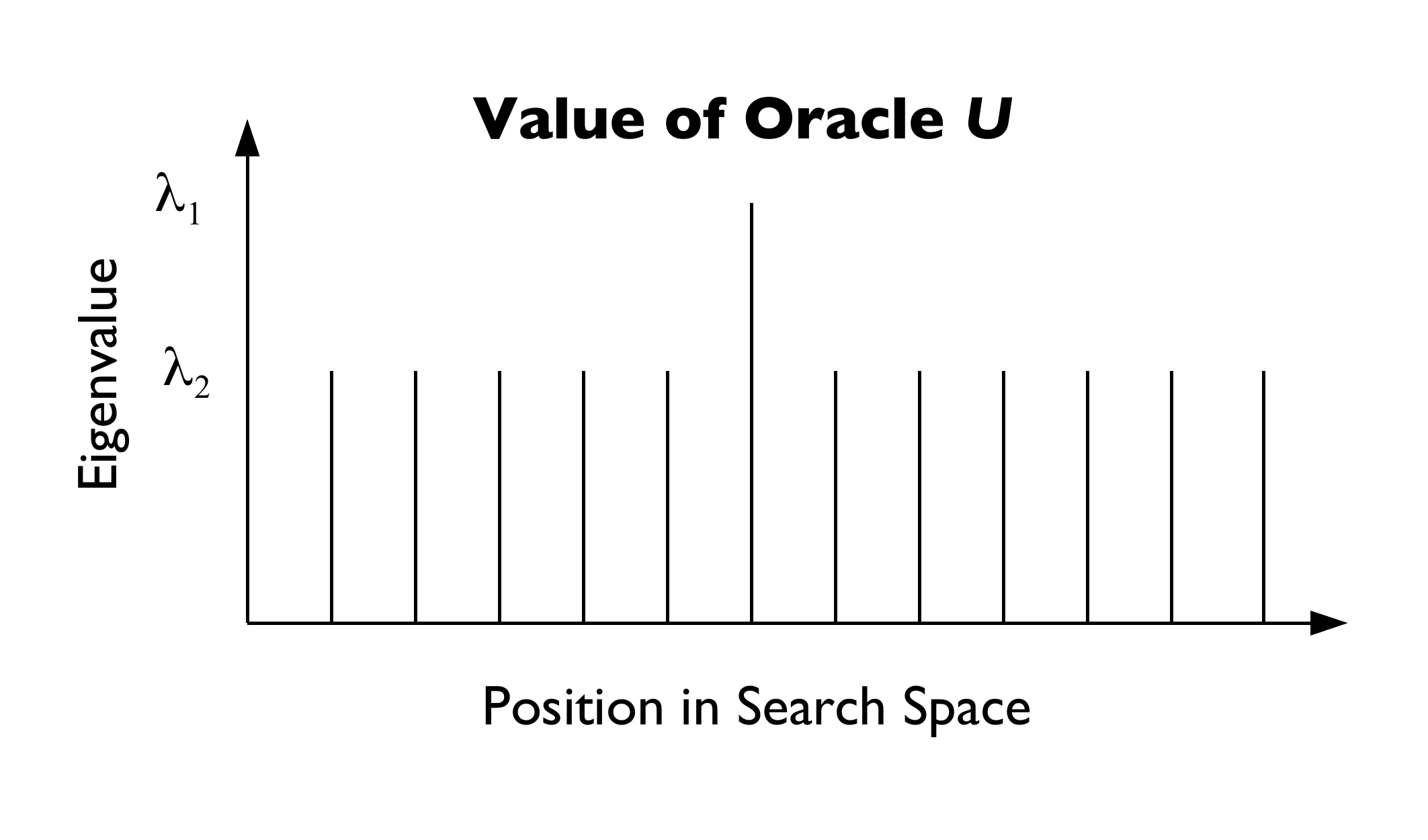}
		\par
	\end{centering}
	\caption{Illustration of the oracles' diagonal entries, assumed to have the unique maximum $\lambda_{1}>0$ and all other eigenvalues of equal amplitude.\label{fig:Illustration}}
\end{figure}

The Appendix provides a formal proof of convergence for IPA continuous global optimization. Here, we focus on discrete optimization for a problem with a single global minimum. We show that the number of IPA steps necessary to amplify the amplitude of the global minimum to a value higher than $1/2=50\%$ scales logarithmically with the number $n\ge 1$ of possible states. The analysis is analogous to the estimation of the number of queries required for amplitude amplification by Grover's algorithm.\cite{Grover.1996.212}
First, we show that IPA converges to the global minimum for the specific case where $U$ is identified with an $n\times n$ diagonal matrix $\mathbf{U}$ with $n$ positive entries $\lambda_{j}>0,\,j=1,\dots,n$ (eigenvalues) with a unique maximum $\lambda_{1}>0$.
For simplicity, we take all other eigenvalues to be $\lambda_{2}$, with
\begin{equation}
	\lambda_{2}<\lambda_{1}.
\end{equation}
Hence, the ``oracle'' $U$ can be expressed as follows
\begin{equation}
	\mathbf{U}
	=\operatorname{diag}\left(\lambda_{2},\ldots,\lambda_{2},\lambda_{1},\lambda_{2},\ldots,\lambda_{2}\right)\in\R^{n\times n},
\end{equation}
where the maximum $\lambda_1$ is the $k$th diagonal entry for some $1\le k\le n$. An illustration is given in Figure~\ref{fig:Illustration}.

We consider a uniform initial density represented by the vector
\begin{equation}
	\mathbf{\rho}_{0}
	=\frac{1}{n}\left(1,\ldots,1\right)\in\R^n.
\end{equation}
The $k$th IPA iteration updates the density distribution as follows
\begin{align}
	\mathbf{\rho}_{k}
	&=\frac{\mathbf{U}\mathbf{\rho}_{k-1}}{\|\mathbf{U}\mathbf{\rho}_{k-1}\|_1}
	=\frac{\mathbf{U}^k\mathbf{\rho}_{0}}{\|\mathbf{U}^k\mathbf{\rho}_{0}\|_1},\\
 	&=\frac{\mathbf{u}_k}{\|\mathbf{u}_k\|_1},
\end{align}
where repeated application of the matrix $\mathbf{U}$ yields
\begin{align}
	\mathbf{u}_{k}
	&=\left(\lambda_{2}^{k},\ldots,\lambda_{2}^{k},\lambda_{1}^{k},\lambda_{2}^{k},\ldots,\lambda_{2}^{k}\right)
\end{align}
with 1-norm ({\em i.e.}, the sum of the absolute values)
\begin{equation}
	\|\mathbf{u}_k\|_1
	=\sum_{j=1}^n|(\mathbf{u}_k)_j|
	=\lambda_1^k+(n-1)\lambda_2^k.
\end{equation}
We note that $\lambda_{1}^{k}>\lambda_{2}^{k}$ since $\lambda_{1}>\lambda_{2}$, so the vector $\mathbf{\rho}_k$ produced after $k$ iterations has only positive entries, a unique maximum
\begin{equation}
	\mathbf{\rho}_{k,\max}
	=\max_{j=1,\dots, n} (\mathbf{\rho}_{k})_j
	=\frac{\lambda_{1}^{k}}{\|\mathbf{u}_k\|_1},
\end{equation}
and all other entries with value
\begin{equation}
	\mathbf{\rho}_{k,\min}
	=\min_{j=1,\dots,n} (\mathbf{\rho}_{k})_j
	=\frac{\lambda_{2}^{k}}{\|\mathbf{u}_k\|_1}.
\end{equation}
Therefore, the minimum to maximum amplitude ratio is
\begin{equation}\label{eq:vmintovmax}
	\frac{\mathbf{\rho}_{k,\min}}{\mathbf{\rho}_{k,\max}}=\left(\frac{\lambda_{2}}{\lambda_{1}}\right)^{k}.
\end{equation}
Each IPA iteration decreases the ratio by a factor of $\lambda_{2}/\lambda_{1}<1$ while the norm is conserved. Therefore, only the maximum entry of the state vector $\mathbf{\rho}_{k}$ survives in the limit of an infinite number of iterations $k\rightarrow+\infty$.
Using the normalization condition,
\begin{equation}\label{eq:normalizationcondition}
	1
	=\|\mathbf{\rho}_{k}\|_1
	=\mathbf{\rho}_{k,\max}+\left(n-1\right)\mathbf{\rho}_{k,\min}
\end{equation}
and inserting the ratio given by Eq.~\eqref{eq:vmintovmax} into the normalization condition introduced by Eq.~\eqref{eq:normalizationcondition}, we can solve for the maximum amplitude $\mathbf{\rho}_{k,\max}$, as follows
\begin{equation}\label{eq:vkmax}
	\mathbf{\rho}_{k,\max}
	=\frac{1}{1+\left(n-1\right)\times\left(\lambda_{2}/\lambda_{1}\right)^{k}},
\end{equation}
which converges to 1 in the limit $k\rightarrow\infty$.

The number of iterations required to amplify the amplitude of the global minimum to a value higher than or equal to $1/2$ is
\begin{equation}
	\frac{1}{1+\left(n-1\right)\times\left(\lambda_{2}/\lambda_{1}\right)^{k}}
	\ge\frac{1}{2}.
\end{equation}
Solving this inequality gives the minimum number of required IPA iterations
\begin{equation}\label{eq:NumberIterations}
	k
	\ge\frac{\log\left(n-1\right)}{\log\left(\lambda_{1}/\lambda_{2}\right)},
\end{equation}
which scales logarithmically with the size of the search space $n\ge 2$ and inverse logarithmically with the ratio of eigenvalues $\lambda_{1}/\lambda_{2}$.

\section{Comparison to Other Methods}\label{sub:power_iteration}
IPA can be compared to the power method\cite{Muntz.1913.43,vMises.1929.58,vMises.1929.152} and
imaginary time propagation.\cite{Metropolis.1949.335,Donsker.1950.551,Anderson.1975.1499,Reynolds.1982.5593} 
The connection between the power method and imaginary time propagation has been discussed,\cite{Greene.2017.4034,Lehtovaara.2007.148,Bader.2013.1304.6845v2,Shani.2017.Analysis,Schwarz.2017.176403} although the relationship between the two methods has yet to be formally analyzed.

We begin with the recurrence relation of the power method. For a matrix $\mathbf{U}\in\C^{n\times n}$ with eigenvalues $\lambda_1,\dots,\lambda_n\in\C$, the subscripts denote the order $|\lambda_{1}|>|\lambda_{2}|\ge...\ge|\lambda_{n}|$.
Given a starting vector $\mathbf{\rho}_{0}\in\C^{n}$ that has a nonzero amplitude along the direction of the eigenvector with the largest eigenvalue $\lambda_1$, the power method produces the following sequence of vectors $\mathbf{\rho}_{k}\in\C^{n}$
\begin{align}\label{eq:power_iteration}
	\mathbf{\rho}_{k}
	=\frac{\mathbf{U}\mathbf{\rho}_{k-1}}{\|\mathbf{U}\mathbf{\rho}_{k-1}\|}
	=\frac{\mathbf{U}^{k}\mathbf{\rho}_{0}}{\|\mathbf{U}^{k}\mathbf{\rho}_{0}\|},
\end{align}
a sequence that converges to an eigenvector associated with the largest eigenvalue $\lambda_1$ independently of the norm $\|\cdot\|$. The resulting convergence is geometric in the ratio\cite{Trefethen:1997aa}
\begin{align}\label{eq:ratepower}
	\left|\frac{\lambda_{2}}{\lambda_{1}}\right|
	<1.
\end{align}
We note that according to the recurrence relation, introduced by Eq.~\eqref{eq:power_iteration}, imaginary time propagation is essentially the power method where $\mathbf{\rho}_0$ represents a trial initial wavefunction in a given basis set and $\mathbf{U}$ is the matrix representation of the Boltzmann operator $e^{-\beta \hat{H}}$, where the Hamiltonian $\hat{H}$ is typically $\hat{H}=\hat{p}^2/(2m)+V$ with $m$ the mass and $\hat{p}=-i\hbar\nabla$ the momentum operator. 

In IPA, however, $\mathbf{\rho}_0({\bf x})$ is a probability density and $U({\bf x})$ can be any integrable, continuous, and strictly positive function of ${\bf x}\in\R^n$ that is maximal at the global minima locations of $V({\bf x})$. As a result, IPA finds the global minima of $V({\bf x})$ while the imaginary time propagation method finds the eigenstate of the Hamiltonian with minimum eigenvalue ({\em i.e.}, the ground state). For the particular choice of $U({\bf x})=e^{-\beta V({\bf x})}$, however, IPA corresponds to the imaginary time propagation with $m=\infty$.

Eq.~\eqref{eq:power_iteration} also shows that IPA differs from the power method because it employs an integrable function $U({\bf x})$ that meets the conditions described in Section~\ref{sub:ipa_sequence} and a probability density function $\mathbf{\rho}_0({\bf x})$ to find the global minima, whereas the power method employs an arbitrary matrix $\mathbf{U}\in\C^{n\times n}$ and a discrete vector $\mathbf{\rho}_0\in\C^n$ to find an eigenvector. This relationship also allows us to use the power method to analyze the convergence rate of IPA for discrete problems, as discussed in the Appendix.

\section{Computational Results}
\begin{figure}[H]
	\begin{centering}
		\includegraphics[angle=0, width=\textwidth]{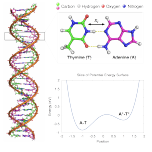} 
		\par
	\end{centering}
	\caption{DNA chain (left) of $D=50$ hydrogen bonds corresponding to 25 hydrogen-bonded adenine-thymine base pairs (inset, top right), with hydrogen bonds shown as dashed yellow lines. Each hydrogen-bonded proton attaches to either base, with energy represented by the double-well potential (bottom right). The global minima positions at $-1$ and $1$ correspond to the depicted A-T form and tautomeric A*-T* form, respectively. Global optimization is performed with $2^8$ possible proton positions for each double well. IPA thus finds the global minimum out of $2^{400}$ possible proton configurations on a PES with $2^{50}$ minima.\label{fig:Chemicals}}
\end{figure}

We demonstrate the capabilities of IPA as applied to the global minimum energy configuration search in a model PES of a DNA chain of $D=50$ hydrogen-bonded adenine-thymine (A-T) base pairs, depicted in Figure~\ref{fig:Chemicals}. The potential energy as a function of the $D$ physical proton coordinates $x_i$ is modeled as a sum of double wells,\cite{Godbeer.2015.13034}
\begin{equation}\label{eq:DNA}
	V(\mathbf{x})
	=\sum_{i=1}^{D=50}0.429~x_i-1.126~x_i^2-0.143~x_i^3+0.563~x_i^4
\end{equation}
parametrized to yield the scaled energy in electronvolts of an A-T/A*-T* base pair as a function of a dimensionless reduced coordinate of a single proton $x_i\in\mathbb{R}$.\cite{Godbeer.2015.13034}
The analytic global minimum $x_i^*=-1$ and local minimum $x_i=1$ correspond to the lowest energy configuration (A-T) and the tautomeric configuration (A*-T*), respectively. Since each proton forms a stable configuration upon adherence to either base, the potential energy surface features $2^{50}$ local minima. Identification of the minimal energy configuration is essential as anomalous hydrogen bonding causes affinity to the incorrect base on replication, which is a proposed mechanism for oncogenesis.\cite{Watson.1953.737,Lowdin.1966.213,Guallar.1999.9922}. Here, global optimization is performed with $d=8$ quantics ({\em i.e.}, in $n=d\times D=400$ dimensions, as the overall dimensionality of the quantics tensor train considered is a product of the number of folding dimensions $d$ and physical dimensions $D$).

\begin{figure}[H]
	\begin{centering}
		\includegraphics[width=0.7\textwidth]{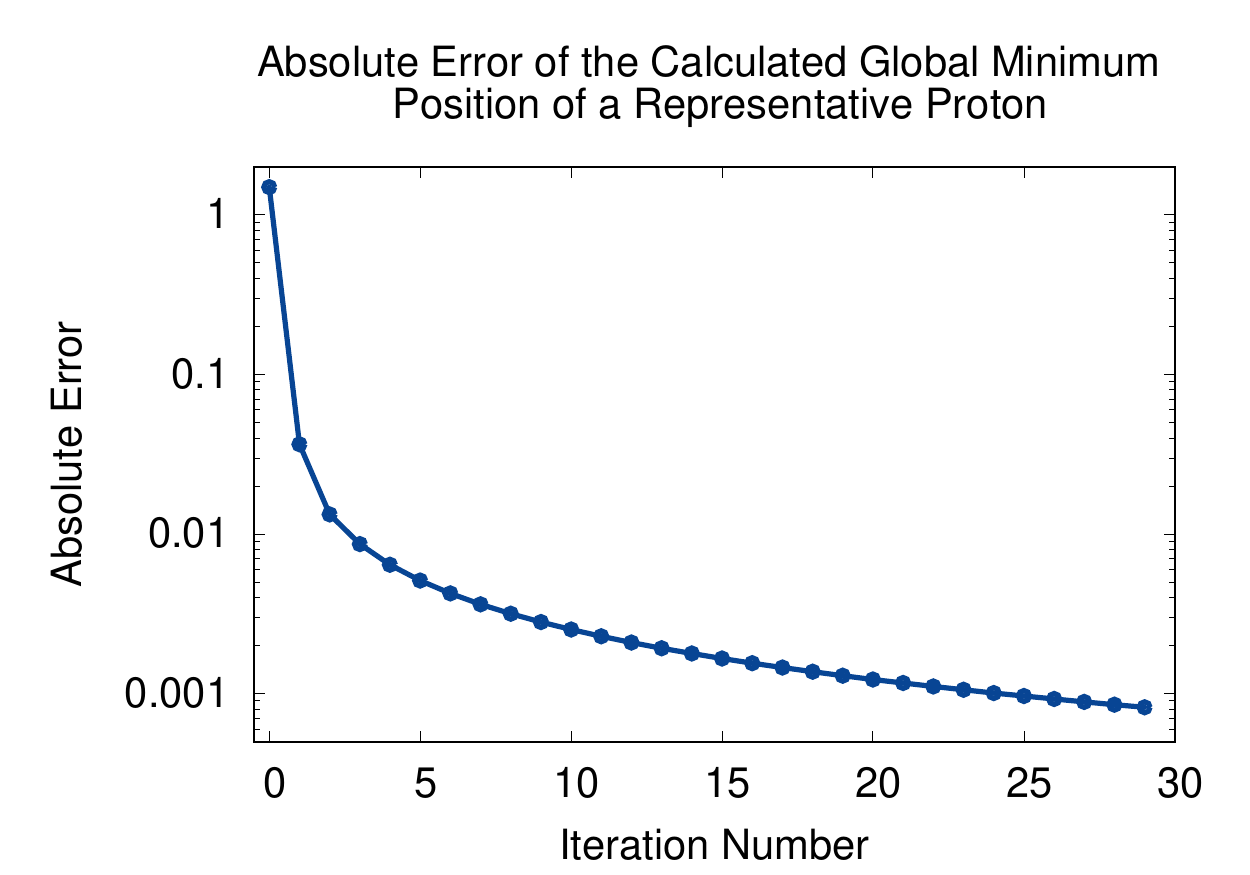} 
		\par
	\end{centering}
	\caption{Expectation value of the position of a representative proton rapidly approaches the known global minimum position of the hydrogen bonding potential, Eq.~\eqref{eq:DNA}.\label{fig:xm0V0}}
\end{figure}

\begin{figure}[H]
	\begin{centering}
		\includegraphics[width=0.7\textwidth]{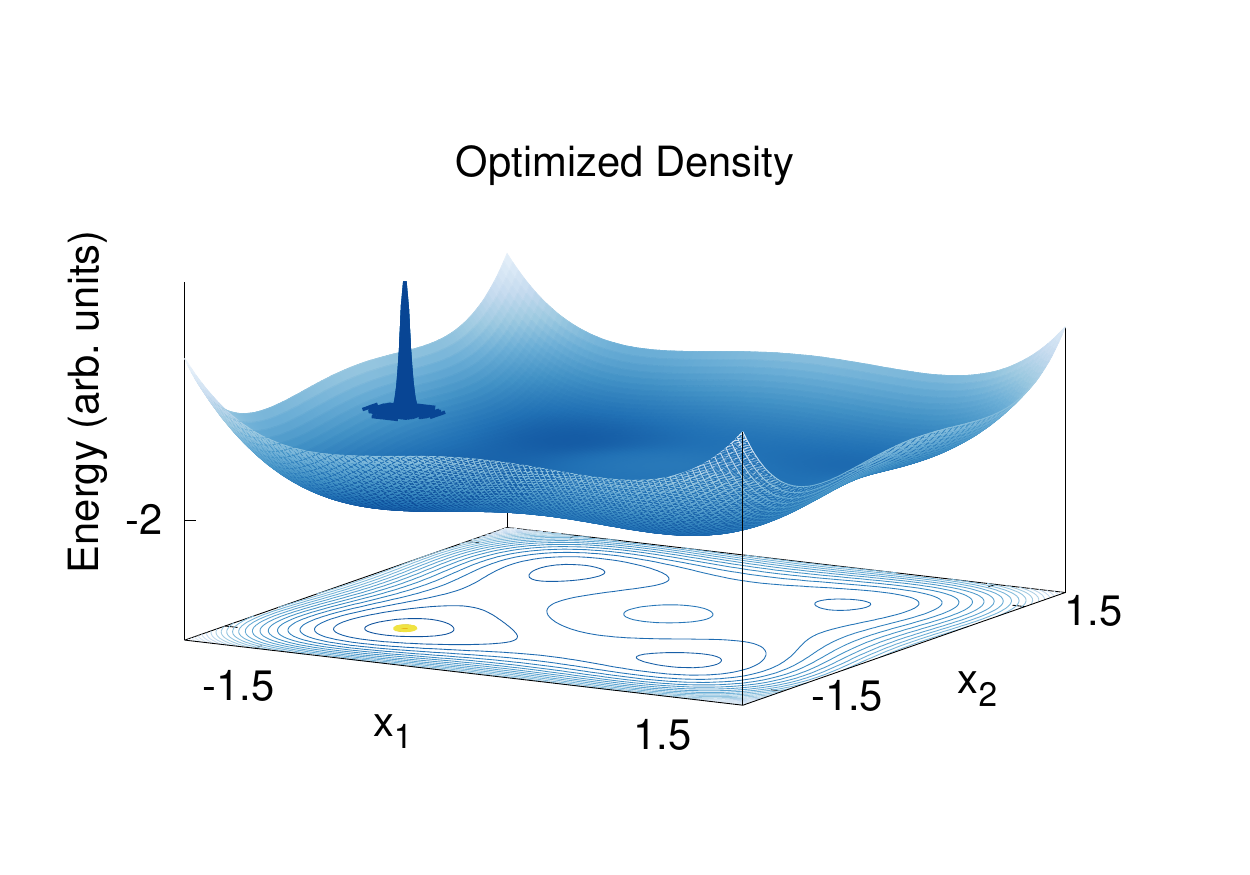} 
		\par
	\end{centering}
	\caption{2-dimensional cut of the 50-dimensional PES (with all other coordinates evaluated at $x=-1$), showing that IPA successfully localizes the final density (dark blue surface) at the global minimum position ${\bf{x}}^\star=(-1,-1) \in\R^2$ (yellow point) of each pair of protons as described by Eq.~\eqref{eq:DNA} (light blue surface). Results are shown for optimization of two hydrogen bonds in the domain $x_1,x_2\in [-1.5, 2.5]$.\label{fig:psiV}}
\end{figure}

IPA correctly identifies the global minimum with the Python code provided in Appendix~\ref{subsub:dna}. As expected, the density initially equally weights all possible proton positions. The expectation value of the position of the initial density lies in the local minimum well, such that gradient descent would not locate the global minimum. IPA iterations successfully concentrate the density at the global minimum well, as evidenced by the rapid convergence of the position expectation value of a representative proton to its global minimum value (see Figure~\ref{fig:xm0V0}). As shown in Figure~\ref{fig:psiV}, after one iteration (with scaling parameter $\beta=10\text{ eV}^{-1}$), the density is localized in the global minimum well; and after $30$ iterations, the density is localized at the global minimum within an absolute error of $10^{-3}$.

\begin{figure}[H]
	\begin{centering}
		\includegraphics[width=\textwidth]{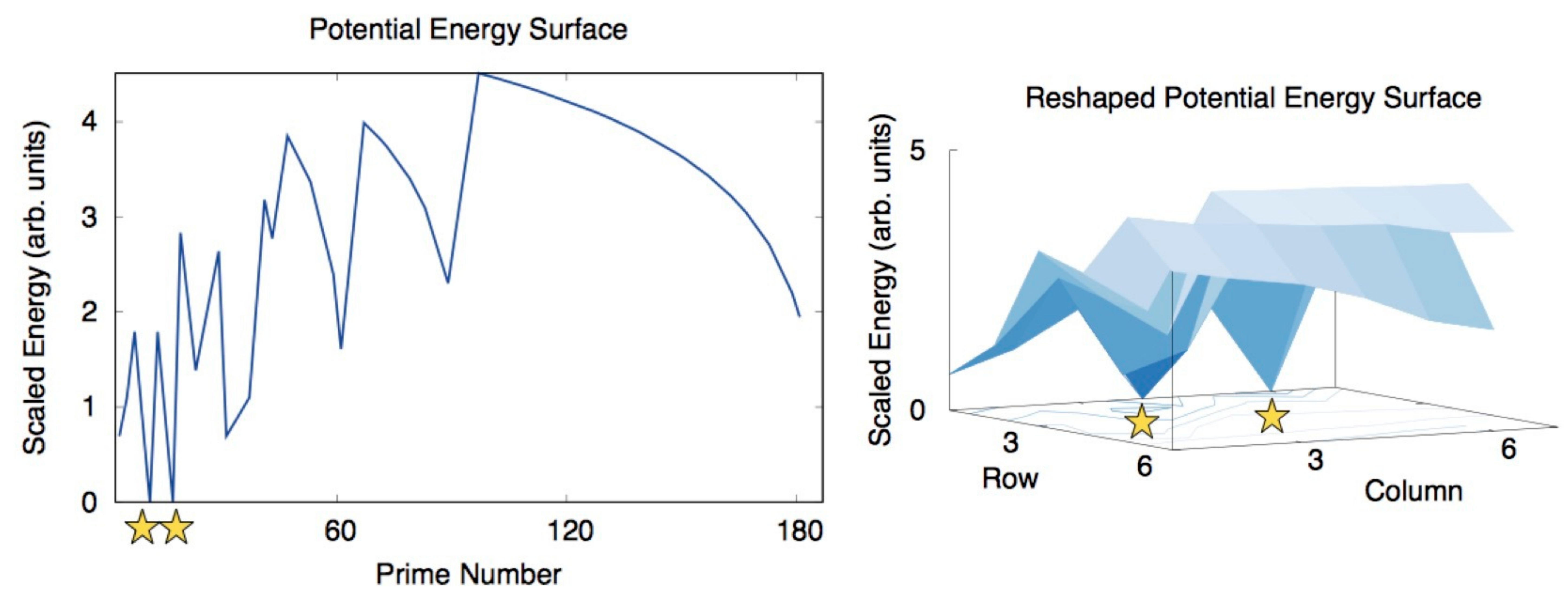} 
		\par
	\end{centering}
	\caption{Scaled potential energy surface $\log(1+V(p))$ for optimization of $V(p)=N\,\operatorname{mod}\,p$, with $N=187$ (left). The process of folding the potential into multidimensional form is illustrated in two reshaping dimensions for a $6 \times 7$ configuration 2-dimensional search space (right). The global minima (starred) correspond to the prime factors of $N=11\cdot 17$. The reported IPA prime factorization of large numbers folds the PES analogously in higher dimensionality $d=3$ to $d=14$.\label{fig:PrimePotential}}
\end{figure}

In addition, this section shows that IPA successfully finds the global minima of the discrete potential 
\begin{equation}\label{eq:ModuloCostFunction}
	\color{black}{V(p)
	=N\,\operatorname{mod}\,p,}
\end{equation}
for $p$ in the set of primes $\{2,3,\dots,p_{\max}\}$, which models a rugged potential energy surface with many local minima and degenerate global minima. This surface enables detection of the prime factors of a given integer $N$, when formulating the factorization problem as a rather challenging global minimum energy search. The modulo operation that defines $V(p)$ in the space of prime numbers $p$ returns the remainder after division of $N$ by $p$. For the difficult problem of optimizing surfaces where the integer $N$ is large (equivalent to prime factorization of large numbers), the Python scripts provided in Appendices~\ref{subsub:multiple_minima} and~\ref{subsub:paired_minima} represent $N$ and operations on $N$ with $3000$-digit precision, using the mpmath library.\cite{Johansson.2013.mpmath} Global optimization of the potential surfaces shows that IPA can resolve the $m$ multiple degenerate prime factors of integers with thousands of digits of the form,
\begin{equation}
	\color{black}{N
	=(p^*_1)^{e_1}\times(p^*_2)^{e_2}\times\cdots\times(p^*_m)^{e_m}},
\end{equation}
where $e_j\ge 1$ is the degeneracy of the prime factor $p_j^*$.
A simple example for $N=187$ is shown in Figure~\ref{fig:PrimePotential}, where the global optima are $p^*_1=11$ and $p^*_2=17$ with $e_1=e_2=1$. In the quantics tensor train (QTT) format employed here, the search space is reshaped such that optimization is performed in $d=3$ to $d=14$ folding dimensions. 

The Python script provided in Appendix~\ref{subsub:multiple_minima} successfully resolves multiple degenerate global minima, regardless of the number of minima, their degeneracy, the distance between minima, or the potential energy barrier separating the minima.

The QTT approximation of $\rho_0\left(p\right)$ provides an accurate and efficient representation of the initial uniform distribution in the search space (the prime numbers $\le N$), folded as a $d$-dimensional $2_1 \times 2_2 \times \cdots \times 2_d$ tensor. The distribution evolves according to the IPA recurrence relation, which increases the amplitude at the global optima while reducing it elsewhere. Application of $U(p)=e^{-\beta V(p)}$ with the scaling parameter $\beta=30$ (arbitrary units) yields a numerically converged final density in only three IPA iterations.

\begin{figure}[H]
	\begin{centering}
		\includegraphics[width=0.7\textwidth]{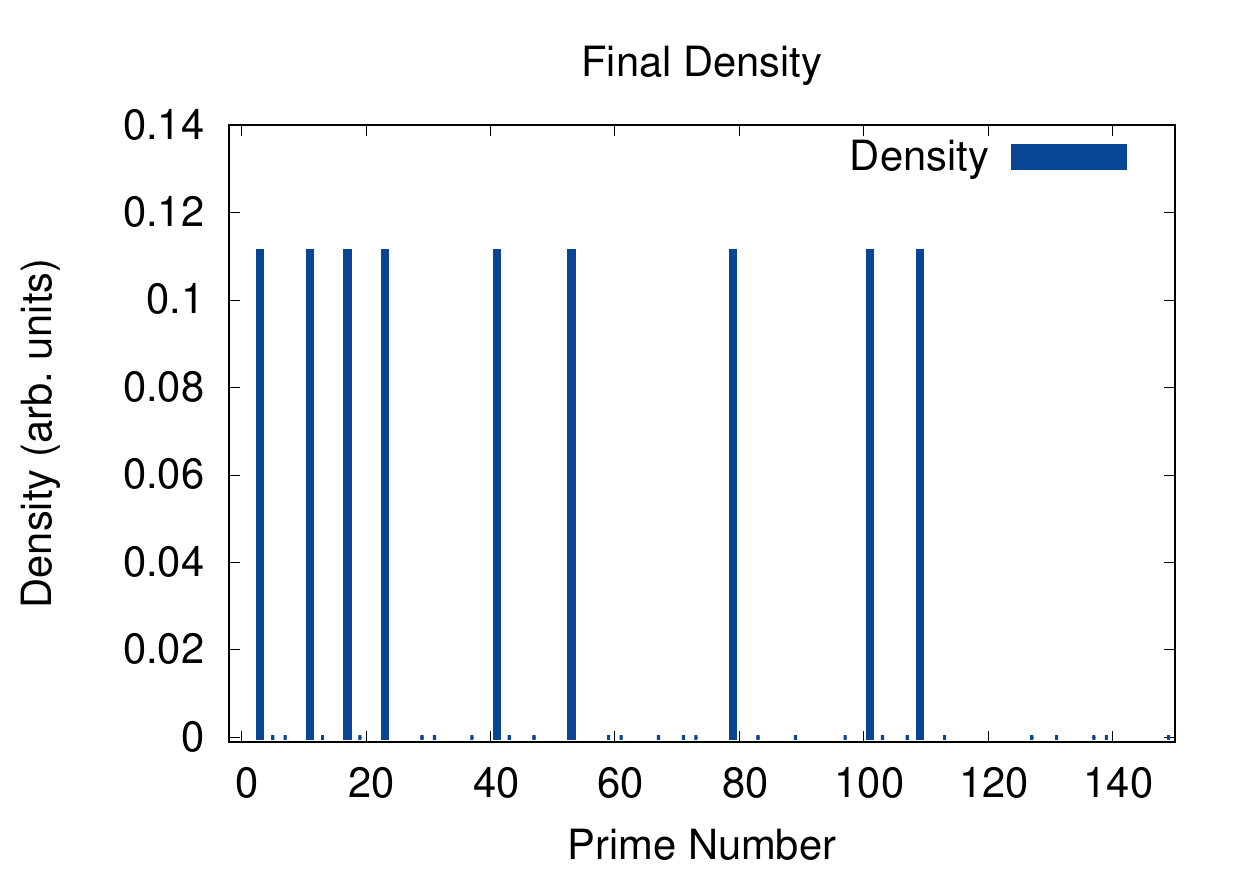} 
		\par
	\end{centering}
	\caption{As expected, the IPA procedure for global optimization of the function Eq.~\eqref{eq:ModuloCostFunction}, for $N$ as defined in the text, yielded a final density in the form of a Dirac comb that was maximal at positions of global optima and zero elsewhere. Only a fraction of the search space is illustrated for clarity.\label{fig:LargestNumber}}
\end{figure}

\begin{figure}[H]
	\begin{centering}
		\includegraphics[width=0.7\textwidth]{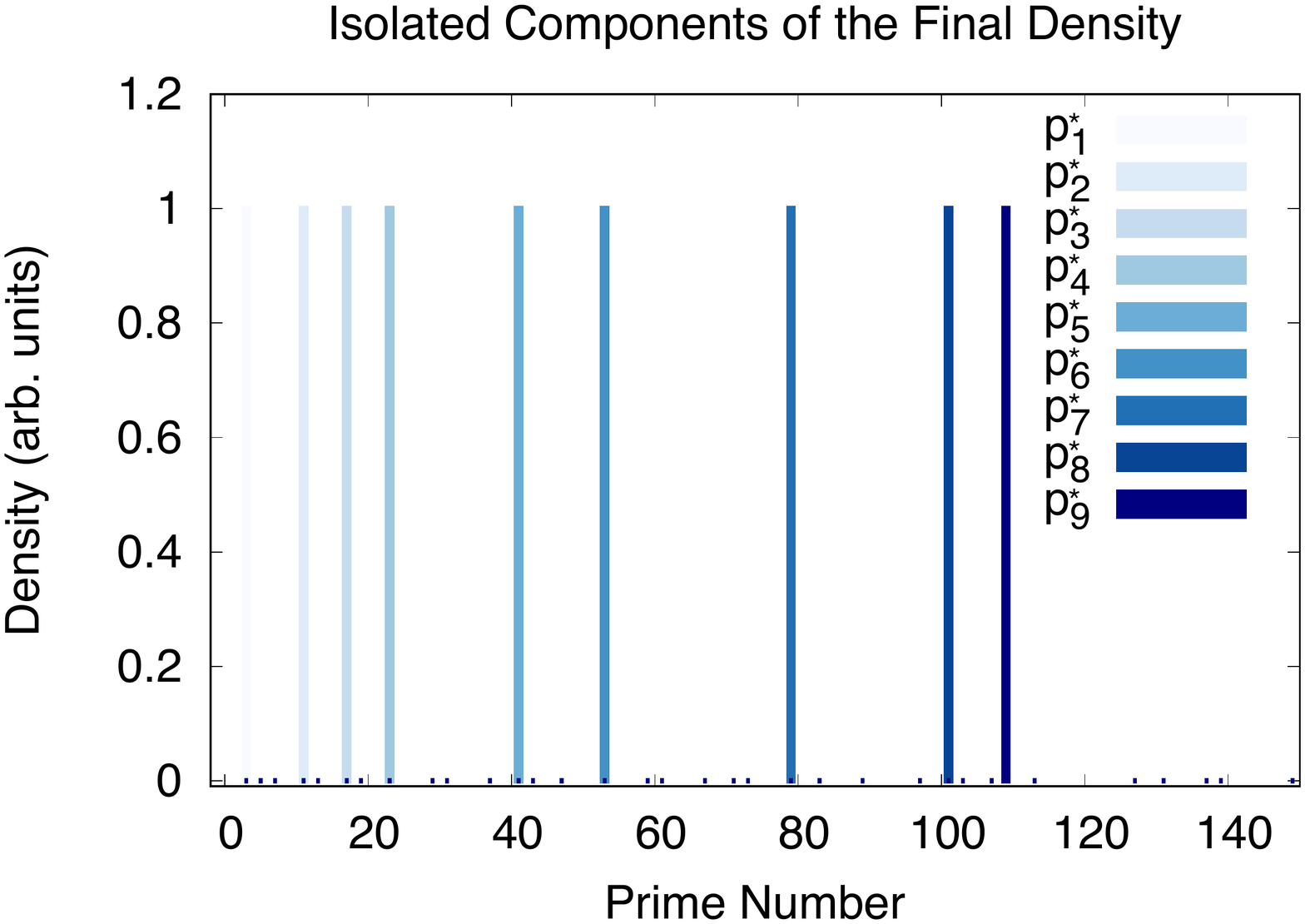} 
		\par
	\end{centering}
	\caption{Dirac delta components of the final density in IPA were successfully isolated without evaluation of the function at all points on the search space via the ramp method for $U=e^{-\tilde\beta\operatorname{ramp}(p)}$ with the parameter $\tilde\beta=0.5$ (arbitrary units).
		The components are found to be located at the global optima of the function Eq.~\eqref{eq:ModuloCostFunction} for the large number $N$.
	Given the size of the search space of prime numbers, the density is shown in a restricted region to enable visualization of its maximal values.\label{fig:IndividualFactors}}
\end{figure}

Figure~\ref{fig:LargestNumber} shows that IPA correctly amplifies the amplitude of the global minima: the degenerate prime factors of $N =(3^2\times 11\times 17\times 23\times 41\times 53\times 79\times 101\times 109)^{200}$, a large integer with 2,773 digits (more than 9,212 bits). Consistent with a Dirac sequence, the final density is maximal for the global minima and nearly zero elsewhere in the search space. Measurement with the ramp function, as described in Section~\ref{sec:measurement}, then successfully resolves the individual global minima as shown in Figure~\ref{fig:IndividualFactors}. IPA thus correctly determines the position of all global minima of the test potential function. 

\begin{figure}[H]
	\begin{centering}
		\includegraphics[width=0.7\textwidth]{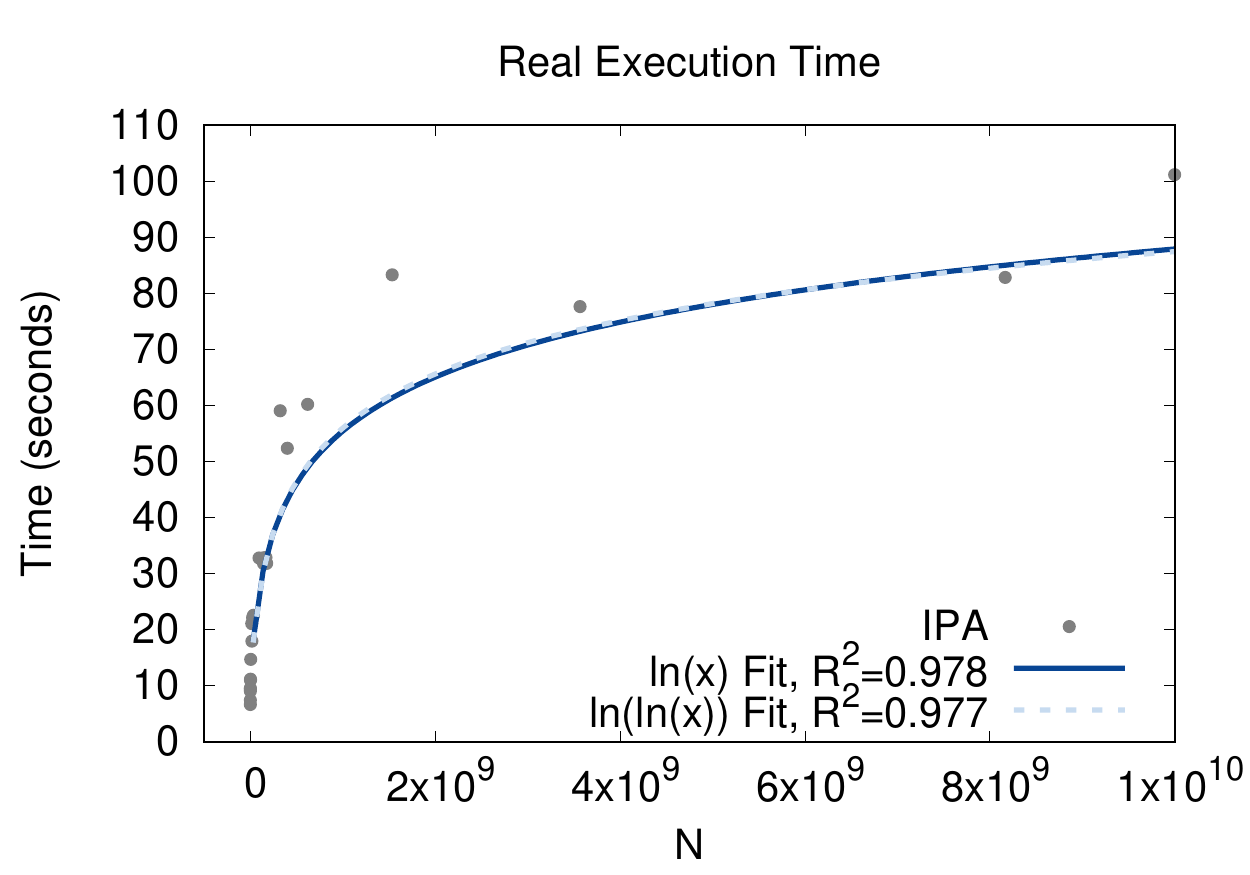} 
		\par
	\end{centering}
	\caption{Real execution time for IPA global optimization of the function
		Eq.~\eqref{eq:ModuloCostFunction} in the twin global minima case ({\em i.e.}, for prime factorization of biprimes), which agrees with the predicted scaling of Section~\ref{sec:ConvergenceRate} and which is comparable to or better than the number of steps required for the rate-limiting part of the foremost quantum approach.\label{fig:Scaling}}
\end{figure}

Figure~\ref{fig:Scaling} shows the IPA execution time as a function of $N$ when the potential Eq.~\eqref{eq:ModuloCostFunction} has two degenerate minima ({\em i.e.}, when solving the factorization of biprimes $N=p_1^*\times p_2^*$ with values up to $9998000099$, where $p_1^*$ and $p_2^*$ are primes.) Results are shown where $U(p)=e^{-\beta V(p)}$ with $\beta=20$ (arbitrary units), which requires only one IPA iteration. The regression analysis shows that the execution time scales approximately as $\mathcal{O}\left(\ln N\right)$ ($R^{2}=0.978$), or $\mathcal{O}\left(\ln\left(\ln N\right)\right)$ ($R^{2}=0.977$).
The logarithmic scaling agrees with the analysis of Section~\ref{sec:ConvergenceRate}, which shows that the resulting scaling for amplitude amplification is comparable to or better than that in optimal quantum search algorithms ({\em e.g.}, Grover quantum search method,\cite{Grover.1996.212} where the number of queries necessary to amplify the amplitude of one out of $N$ possible states scales as $\mathcal{O}(\sqrt{N})$).

\section{Discussion}
The QTT implementation of IPA illustrates the possibility of developing efficient algorithms for classical computing and chemistry. Analogous to quantum computing algorithms, superposition states can be evolved by applying a sequence of unitary transformations, and the outcome of the calculation corresponds to a ``measurement'' ({\em i.e.}, an expectation value obtained with the evolved superposition). The QTT representation avoids the curse of dimensionality, enabling benchmark calculations that would be otherwise impossible on classical high-performance computing facilities. We find that such a computational strategy enables IPA to perform quite efficiently, bypassing the usual limitations of traditional optimization methods. Therefore, it is natural to anticipate that IPA should be of great interest for a wide range of applications, including optimization problems in molecular and electronic structure calculations.

\section*{Acknowledgements}
The authors are grateful for conversations with Dr.~Erik T.~J.~Nibbering, Dr.~Caroline Lasser, and Dr.~Maximilian Engel and thank Krystle Reiss for generation of molecular structure images. M.~B.~S.~acknowledges financial support from the Yale Quantum Institute Postdoctoral Fellowship, the National Science Foundation Graduate Research Fellowship grant number DGE-1144152, and the Blue Waters Graduate Research Fellowship, part of the Blue Waters sustained-petascale computing project supported by the National Science Foundation (awards OCI-0725070 and ACI-1238993) and the State of Illinois. Blue Waters is a joint effort of the University
of Illinois at Urbana-Champaign and its National Center for Supercomputing Applications. V.S.B. acknowledges support from the NSF Grant No. CHE-1900160 and high performance computing time from NERSC and the Yale High Performance Computing Center.

\section*{TOC Graphic}
\begin{centering}
	\includegraphics[width=0.7\textwidth]{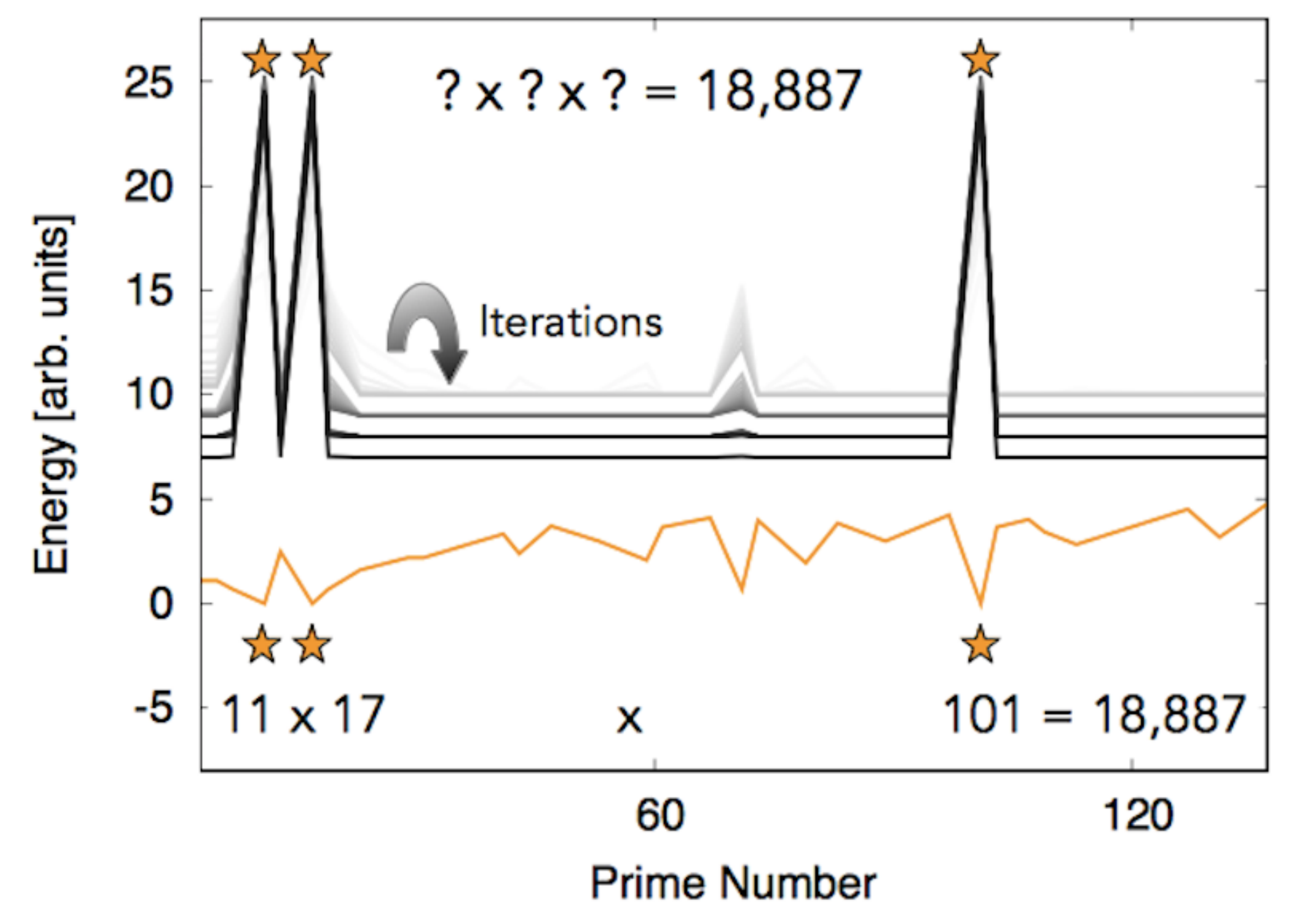}
	\par
\end{centering}
%

\section*{Keywords}
Global optimization, quantum computing, tensor networks, prime factorization, quantum superposition

\appendix
\part*{Appendix}

\section{Proof of Convergence}\label{subsub:proof_convergence}
This section shows that the sequence generated by the IPA recurrence relation converges to a delta distribution $\delta(x-x^*)$ if $V(x)$ has a single global minimum at $x=x^*$. An analogous proof can be provided for surfaces with multiple global minima by generalization of the concept of a Dirac sequence.

The sequence of densities $\rho_k(x)$ converges to the delta distribution as a Dirac sequence:
\begin{enumerate}
	\item[(i)] For all $k\in\N$ and all $x\in\R$: $\rho_{k}(x)\ge0$;
	\item[(ii)] For all $k\in\N$: $\rho_k\in L^1(\R)$ and $\displaystyle\int_{\R}\mathrm{d}x\,\rho_{k}(x)=1$;
	\item[(iii)] For all $\e>0$: $\displaystyle\lim_{k\to\infty}\int_{\R \setminus (x^*-\e,x^*+\e)}\mathrm{d}x\,\rho_{k}(x)=0$, where the integral is evaluated over the real line except the interval $(x^*-\e,x^*+\e)$;
\end{enumerate}
These conditions guarantee the area under the curve $\rho_k(x)$ is concentrated near the global minimum location $x^*$, provided the number of iterations $k$ is sufficiently large.

\begin{figure*}
	\includegraphics[width=80mm]{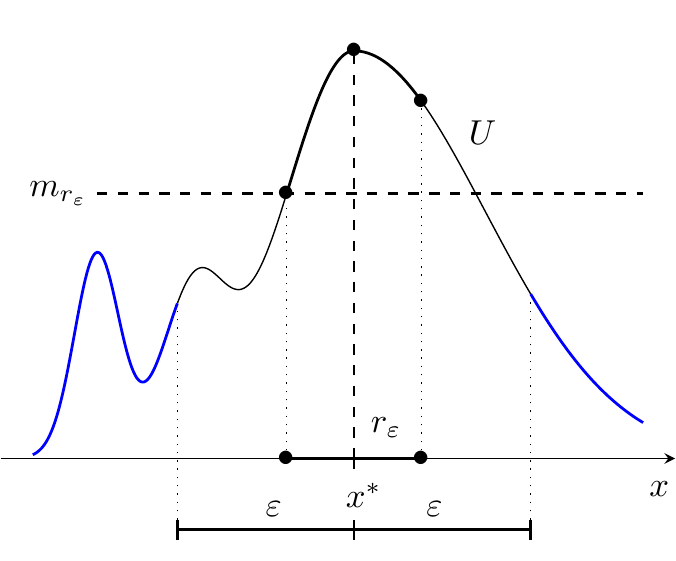}
	\caption{There exists a radius $r_\e>0$ such that the minimum $m_{r_\e}$ on $[x^*-r_\e,x^*+r_\e]$ is a strict upper bound (dashed line) for all values outside the interval $(x^*-\e,x^*+\e)$ (shown in blue), since $U(x)$ is a continuous function with a single global maximum at $x=x^*$.\label{fig:1}}
\end{figure*}

The properties (i) and (ii) follow by construction of the IPA sequence. To prove property (iii), let $\e>0$ be a positive distance.
For a radius $r>0$, we denote the minimum of $U(x)$ on the interval $[x^*-r,x^*+r]$ by
\begin{align}\label{eq:m_eps}
	m_r
	=\min_{x\in[x^*-r,x^*+r]} U(x).
\end{align}
Since by assumption $U(x)$ is continuous with a single global maximum at $x=x^*$, there exists a radius $r_\e>0$ such that the number $m_{r_\e}$ is a positive and strict upper bound for $U(x)$ outside the interval $(x^*-\e,x^*+\e)$, as follows (cf. Figure~\ref{fig:1}):
\begin{align}\label{proof:seperated}
	\frac{U(x)}{m_{r_\e}}
	<1\quad
	\text{for all $x\in\R\setminus(x^*-\e,x^*+\e)$.}
\end{align}
%
%

We then introduce the probability
\begin{align}
	p_{\e}
	=\int_{x^*-r_\e}^{x^*+r_\e}\mathrm{d}x\,\rho_0(x)
	>0,
\end{align}
and according to the definition of the minimum $m_{r_\e}>0$, introduced by Eq.~\eqref{eq:m_eps}, for all $k\ge 1$ we obtain the norm,
\begin{align}
	\|U^{k}\rho_0\|_{L^{1}}
	=\int_{\R}\mathrm{d}x\,U(x)^{k}\rho_0(x)
	\ge m_{r_\e}^{k}\int_{x^*-r_\e}^{x^*+r_\e}\mathrm{d}x\,\rho_0(x)
	=m_{r_\e}^{k}p_{\e},
\end{align}
which gives the bound
\begin{align}
	\rho_{k}(x)
	=\frac{U(x)^k\rho_0(x)}{\|U^{k}\rho_0\|_{L^{1}}}
	\le\frac{\|\rho_0\|_{\infty}}{p_{\e}}\left(\frac{U(x)}{m_{r_\e}}\right)^{k}
	\quad\text{for all $x\in\R$},
\end{align}
where $\|\rho_0\|_{\infty}$ is the supremum $\sup_{x\in\R}|\rho_0(x)|$.
According to Eq.~\eqref{proof:seperated}, $U(x)/m_{r_\e}<1$ for all positions outside the interval $(x^*-\e,x^*+\e)$.
Hence, we conclude that the density after $k$ iterations is bounded for all those positions $x$ and all iterations $k\ge 1$, as follows:
\begin{align}
	\rho_k(x)
	\le\frac{\|\rho_0\|_{\infty}}{p_{\e}}\frac{U(x)}{m_{r_\e}},
\end{align}
showing that the sequence is dominated by an integrable function. Thus, the Lebesgue-dominated convergence theorem yields
\begin{align}
	\lim_{k\to\infty}\int_{\R\setminus(x^*-\e,x^*+\e)}\mathrm{d}x\,\rho_k(x)
	=\int_{\R\setminus (x^*-\e,x^*+\e)}\mathrm{d}x\,\lim_{k\to\infty}\rho_k(x)
	=0.
\end{align}
%

\section{Power Method: Convergence Rate Analysis}\label{sub:AltConvergence}
We consider a diagonal matrix $\mathbf{U}\in\R^{n\times n}$ whose entries are given by the values of $U(x)$ at the equally spaced positions $a=x_1<x_2<...<x_n=b$ with $\Delta x=x_{j+1}-x_j=(b-a)/(n-1)$ in the finite interval $[a,b]$, that is,
\begin{align}\label{eq:def_U2}
	\mathbf{U}
	=\operatorname{diag}\big(U(x_{1}),U(x_{2}),\dots,U(x_{n})\big).
\end{align}
We consider an initial vector whose entries are given by the values of the initial density $\rho_0(x)$ at the same positions,
\begin{align}
	\mathbf{\rho}_0
	=\big(\rho_0(x_{1}),\rho_0(x_{2}),\dots,\rho_0(x_{n})\big)
	\in\R^n.
\end{align}
When $n$ is sufficiently large, we obtain the following approximation for all iterations:
\begin{align}\label{eq:approx_pi_ipa}
	\|\mathbf{U}^{k}\mathbf{\rho}_0\|_{1}
	=\sum_{j=1}^{n}U(x_{j})^{k}\rho_0(x_{j})
	\approx\frac{1}{\Delta x}\int_{\R}\mathrm{d}x\,U(x)^{k}\rho_0(x)
	=\frac{1}{\Delta x}\|U^{k}\rho_0\|_{L^{1}}.
\end{align}
In the following, we denote by $\mathbf{\rho}^*\in\R^n$ the vector whose $j$th coordinate equals 1 if $U(x_j)=\lambda_1$ is the dominant eigenvalue of $\mathbf{U}$ and zero otherwise.
Moreover, we introduce the constant
\begin{align}
	c
	=\frac{1}{\#\{j\mid U(x_j)=\lambda_1\}},
\end{align}
where we use the notation $\#A$ for the cardinality {\em (i.e.}, the number of elements in the set).
The definition of $\mathbf{U}$ in Eq.~\eqref{eq:def_U2} yields that the sequence $\mathbf{\rho}_{1},\mathbf{\rho}_{2},\dots$ produced by the power iteration ({\em i.e.}, Eq.~\eqref{eq:power_iteration} using the norm $\|\cdot\|_1$) converges to $c\times\mathbf{\rho}^*$ if $\mathbf{\rho}^*$ is the uniform distribution.
Using the approximation in Eq.~\eqref{eq:approx_pi_ipa}, we conclude that the density $\rho_k$ produced by IPA can be approximated at a given grid point $x_j$ as
\begin{align}
	\rho_k(x_j)
	=\frac{U(x_{j})^{k}\rho_0(x_{j})}{\|U^{k}\rho_0\|_{L^{1}}}
	\approx\frac{1}{\Delta x}\frac{(\mathbf{U}^{k}\mathbf{\rho}_0)_{j}}{\|\mathbf{U}^{k}\mathbf{\rho}_0\|_{1}}
	\overset{k\to\infty}{\longrightarrow}\frac{c}{\Delta x}\mathbf{\rho}^*_j.
\end{align}
In the special case where $\mathbf{U}$ has a single unique dominant eigenvalue ({\em i.e.} $\lambda_1=U(x_l)$ for some unique $l\in\{1,\dots,n\}$), we get $\mathbf{\rho}^*_j$ is the Kronecker delta $\delta_{j,l}$.
This allows us to confirm that IPA generates a Dirac sequence at the global minimum for discrete optimization problems. The relationship of this expression to that of the power method also shows that IPA inherits the geometric convergence rate in the ratio $\lambda_2/\lambda_1<1$ from the power method, in agreement with the alternative analysis introduced in Section~\ref{sec:ConvergenceRate}.

To further specify the convergence rate of IPA, we relate the ratio $\lambda_2/\lambda_1$ to the grid size $\Delta x>0$ in IPA. This is accomplished by classifying the steepness of $U(x)$ around its maximum location $x^*$ via local approximations by polynomials of even degree.
If there exist parameters $\alpha>0$ and $\gamma\ge 1$ such that
\begin{align}\label{eq:bound_U}
	U(x)
	\ge U(x^{*})-\alpha(x-x^{*})^{2\gamma}
\end{align}
for all $x\in (x^*-\Delta x,x^*+\Delta x)$, then the eigenvalue $\lambda_{2}$ is bounded from below by $U(x^{*})-\alpha\Delta x^{2\gamma}$. Therefore, we conclude that the rate of convergence is bounded as
\begin{align}
	\frac{\lambda_{2}}{\lambda_{1}}
	\ge\frac{U(x^{*})-\alpha\Delta x^{2\gamma}}{U(x^{*})}
	=1-\frac{\alpha}{U(x^{*})}\Delta x^{2\gamma}.
\end{align}
In particular, $\lambda_2/\lambda_1 \to 1$ as $\Delta x\to 0$.

\section{Global Minimum Energy Configuration of Hydrogen Bonds}\label{subsub:dna}
The following Python script illustrates the use of IPA to find the global minimum energy configuration of 50 adenine-thymine (A-T) hydrogen bonds in a DNA chain with the ttpy library installed from http://github.com/oseledets/ttpy.
 
{\small{\singlespacing
\begin{lstlisting}
import numpy as np
from mpl_toolkits.mplot3d import Axes3D
from matplotlib import cm
import matplotlib.pyplot as plt
import tt

beta=10
dim=2
eps=1.0e-14
rma=3
nsteps=30
d=8
npts=2**d
xmin=-1.5
xmax=2.5
dx=(xmax-xmin)/npts

def gen_1d(mat,e,i,d):
    w=mat
    for j in range(i):
        w=tt.kron(e,w)
    for j in range(d-i-1):
        w=tt.kron(w,e)
    return w

def rhoo(input):
    out=1.+0.*np.sum(input,axis=1)
    return(out);

def V(input):
    out=np.sum(0.429*input-1.126*input**2-0.143*input**3+0.563*input**4,axis=1)
    return(out);

def plot_f2d(x,rho,V,t,vm):
    fig=plt.figure()
    yy,xx=np.meshgrid(x, x)
    ax=fig.add_subplot(211, projection='3d')
    ax.plot_surface(xx,yy,rho+10,cmap=cm.coolwarm,antialiased=False)
    ax.plot_surface(xx,yy,V,cmap=cm.coolwarm,antialiased=False)
    ax.set_xlabel("x (arb. units)")
    ax.set_ylabel("y (arb. units)")
    ax.set_zlabel("Energy (arb. units)")
    ax.set_title("Pointer State")
    ax1=plt.subplot(212)
    plt.plot(t[:j],vm[:j],'o-')
    plt.xlabel("Iteration Number")
    plt.ylabel("<V> (arb. units)")
    ax1.set_title("Expectation Value of Potential")
    plt.show()

def plot_f(t,vm):
    fig = plt.figure()
    plt.plot(t[:j],vm[:j],'o-')
    plt.xlabel("Iteration Number")
    plt.ylabel("<V> (arb. units.")
    plt.title("Expectation Value of Potential")
    plt.show()

if __name__ == "__main__":
    x=np.arange(xmin, xmax, dx)
    t=np.arange(0,nsteps,1)
    xm=np.zeros((nsteps,dim))
    vm=np.zeros(nsteps)
    ttone=tt.ones(2, d)
    xx=np.reshape(x, [2]*d)
    xx=tt.tensor(xx,eps)
    ttoned=ttone
    for ii in range(dim-1):
      ttoned=tt.kron(ttone,ttoned)
    txx=[gen_1d(xx,ttone,i,dim) for i in range(dim)]
    ttrho=tt.multifuncrs2(txx, rhoo, eps,verb=0,rmax=rma)
    norm=tt.dot(ttrho,ttoned)
    ttrho=ttrho*(1.0/norm)
    
    ttVV=tt.multifuncrs(txx,V,eps,verb=0,rmax=rma)

    for j in range(nsteps):
        vm[j]=tt.dot(ttrho,ttVV)
        for k in range(dim):
            xm[j,k]=tt.dot(ttrho,txx[0])
        print("j=,Vm=",j,vm[j])
        if j == nsteps-1:
            if dim == 2:
                ttrhor=np.reshape(ttrho.full(),[2**d]*dim)
                ttVVr =np.reshape(ttVV.full(),[2**d]*dim)
                plot_f2d(x,ttrhor,ttVVr,t,vm)
            else:
                plot_f(t,vm)
        tto = lambda r:np.exp(-beta*V(r))
        ttoracle=tt.multifuncrs2(txx,tto,eps,verb=0,rmax=rma)
        ttrho=ttrho*ttoracle
        ttrho=ttrho.round(eps,rma)
        norm=tt.dot(ttrho,ttoned)
        ttrho=ttrho*(1.0/norm)

    for k in range(dim):
        print("j=,k=,xm=,Vm=",j,k,xm[j,k],vm[j])
\end{lstlisting}}}

\section{Multiple Degenerate Global Minima}\label{subsub:multiple_minima}
The following Python script illustrates the implementation of IPA as applied to finding multiple degenerate global minima corresponding to the degenerate prime factors of the integer $N =(3^2\times 11\times 17\times 23\times 41\times 53\times 79\times 101\times 109)^{200}$ with 2,773 digits, when using the ttpy library installed from http://github.com/oseledets/ttpy.
 
{\small{\singlespacing
\begin{lstlisting}
import numpy as np
from numpy import zeros,reshape,sqrt,arange,vectorize,extract,int

import matplotlib.pyplot as plt
import tt

import mpmath
from mpmath import mp,mpf,floor,exp,nint

def parameters():
    global dim,eps,num,rmax,nsteps,d,searchspacesize,beta,betaprime
    num=mpf(3*3*11*17*23*41*53*79*101*109)**200
    beta=30
    betaprime=0.5
    dim=1
    eps=1.0e-100
    rmax=100
    nsteps=3
    d=6
    searchspacesize=2**d
    return()

def rhoo(input):
    V=1.0+0*input
    return V

def is_prime(n):
    if n % 2 == 0 and n > 1: 
        return False
    return all(n % i for i in range(3,int(sqrt(n))+1,2))

def tto(input, param=None):
    global num,beta
    nevals,dim=input.shape
    out=np.zeros((nevals,))
    for ii in range(nevals):
        a=num-nint(input[ii,0])*floor(num/nint(input[ii,0]))
        out[ii]=input[ii,1]*exp(-beta*a)
    return(out)
    
def newtto(input, param=None):
    global num,betaprime
    nevals,dim=input.shape
    out=np.zeros((nevals,))
    for ii in range(nevals):
        a=input[ii,0]
        out[ii]=input[ii,1]*exp(-betaprime*a)
    return(out)

def ttround(input, param=None):
    nevals, dim=input.shape
    out=np.zeros((nevals,))
    for ii in range(nevals):
        out[ii]=np.round(input[ii,1])
    return(out)
    
def ttremovefactor(input, param=None):
    global largestfactor
    nevals,dim=input.shape
    out=np.zeros((nevals,))
    for ii in range(nevals):
        if input[ii,0]-0.5 < largestfactor:
            out[ii]=0.
        else:
            out[ii]=np.round(input[ii,1])
    return(out)

if __name__ == "__main__":
    global eps,num,rmax,nsteps,d,searchspacesize,largestfactor
    np.random.seed(1234)
    mp.dps = 3000
    parameters()
    print(num)
    a=arange(2, 10**6)
    foo=vectorize(is_prime)
    pbools=foo(a)
    primes=extract(pbools, a)
    pp=np.zeros(searchspacesize,dtype=float)
    for j in range(searchspacesize):
        pp[j]=primes[j]
    ttpp=tt.tensor(reshape(pp,[2]*d))
    lprimes=[]
    ttrho=tt.multifuncrs2([ttpp],rhoo,eps,verb=0,rmax=rmax)
    for k in range(nsteps):
        ttrho=tt.multifuncrs([ttpp,ttrho],tto,eps,verb=0,
            rmax=rmax)
    ttrhostore=tt.multifuncrs([ttpp,ttrho],ttround,eps,verb=0,
        rmax=rmax)
    ttrho=ttrho*(1.0/(ttrho.norm())**2)
    plt.bar(pp,reshape(ttrho.full(),searchspacesize),color='red',ls='-',
        label='ttrho')
    plt.xlabel("Prime Number")
    plt.ylabel("Density [arb. units.]")
    plt.title("Optimized Density")
    plt.xlim(1,128)
    plt.pause(1.0)
    plt.savefig('diraccomb.png')
    
    largestfactor=0.
    count=0
    while num > 1:
        count=count+1
        ttrho=tt.multifuncrs([ttpp,ttrhostore],ttremovefactor,eps,verb=0,
            rmax=rmax)
        for k in range(nsteps):
            ttrho=tt.multifuncrs([ttpp,ttrho],newtto,eps,verb=0,
              rmax=rmax)
        ttrho=ttrho*(1.0/ttrho.norm())
        ev=nint(tt.dot(ttpp,ttrho))
        largestfactor=ev
        lprimes.append(ev)
        num=num/ev
        while nint(num)%nint(ev) == 0:
            lprimes.append(ev)
            num=num/ev
        plt.clf()
        plt.bar(pp,reshape(ttrho.full(),searchspacesize),color='red',ls='-',
          label='Density')
        plt.xlabel("Prime Number")
        plt.ylabel("Density [arb. units.]")
        plt.title("Prime Factor %i" % count)
        print("prime factors=",lprimes,num)
        plt.xlim(1,128)
        plt.savefig('primefactor'+str(count)+'.png')
\end{lstlisting}}}

\section{Paired Degenerate Global Minima}\label{subsub:paired_minima}

The following Python script illustrates the implementation of IPA as applied to finding the prime factors of the biprime $N=99\,989\times 99\,991$ by resolving the degenerate global minima of the mod function, as described in the text, while using the ttpy library installed from http://github.com/oseledets/ttpy.

{\small{\singlespacing
\begin{lstlisting}
import numpy as np
from numpy import zeros,reshape,sqrt,arange,vectorize,extract,int,empty_like
import tt
import mpmath
from mpmath import mp,mpf,floor,exp,nint

def parameters():
    global dim,eps,num,rmax,nsteps,d,searchspacesize,beta
    num=mpf(99989*99991)
    sqrtnum=sqrt(num)
    if sqrtnum < 24:
        d=3
    elif sqrtnum < 60:
        d=4
    elif sqrtnum < 138:
        d=5
    elif sqrtnum < 314:
        d=6
    elif sqrtnum < 728:
        d=7
    elif sqrtnum < 1622:
        d=8
    elif sqrtnum  < 3674:
        d=9
    elif sqrtnum < 8168:
        d=10
    elif sqrtnum < 17882:
        d=11
    elif sqrtnum < 38892:
        d=12
    elif sqrtnum < 84048:
        d=13
    elif sqrtnum < 180512:
        d=14
    else:
        print("Error: Dimension not implemented.")
        quit()
    rmax=100
    beta=20
    dim=1
    eps=1.0e-100
    nsteps=1
    searchspacesize=2**d
    return()

def rhoo(input):
    V=1.0+0*input
    return V
    
def is_prime(n):
    if n % 2 == 0 and n > 1: 
        return False
    return all(n % i for i in range(3,int(sqrt(n))+1,2))

def tto(input, param=None):
    global num,beta
    nevals,dim=input.shape
    out=np.zeros((nevals,))
    for ii in range(nevals):
        a=num-nint(input[ii,0])*floor(num/nint(input[ii,0]))
        if a > 10:
          a=10
        out[ii]=input[ii,1]*exp(-beta*a)
    return(out)

if __name__ == "__main__":
    global eps,num,rmax,nsteps,d,searchspacesize,ttavg
    np.random.seed(1234)
    mp.dps=2000
    parameters()
    print(num)
    a=arange(2, 10**6)
    foo=vectorize(is_prime)
    pbools=foo(a)
    primes=extract(pbools, a)
    pp=np.zeros(searchspacesize,dtype=float)
    for j in range(searchspacesize):
        pp[j]=primes[j]
    ttpp=tt.tensor(reshape(pp,[2]*d))
    lprimes=[]
    ttrho=tt.multifuncrs2([ttpp],rhoo,eps,verb=0,rmax=rmax)
    for k in range(nsteps):
        ttrho=tt.multifuncrs([ttpp,ttrho],tto,eps,verb=0,rmax=rmax)
        ttrho=ttrho*(1.0/(ttrho.norm())**2)
    ttrhostore=ttrho
    ttavg=nint(tt.dot(ttpp,ttrho))
    heaviside=empty_like(pp)
    for j in range(searchspacesize):
        if pp[j]-0.5 > ttavg:
            heaviside[j]=0.
        else:
            heaviside[j]=1.
    ttheaviside=tt.tensor(reshape(heaviside,[2]*d))
    ttrho=ttheaviside*ttrhostore
    ev=nint(tt.dot(ttpp,ttrho*(1.0/(ttrho.norm()))))
    num=nint(num/ev)
    lprimes.append(ev)
    lprimes.append(num)
    print("prime factors=",lprimes,num)
\end{lstlisting}}}

\bibliography{PrimeFactorization}
\end{document}